\documentclass[fleqn,twoside]{article}
\usepackage{espcrc2}

\usepackage{graphicx}
\usepackage{epsfig}

\usepackage[figuresright]{rotating}

\newcommand{\AmS}{{\protect\the\textfont2
  A\kern-.1667em\lower.5ex\hbox{M}\kern-.125emS}}

\title{A lattice study of $\Lambda_b$ semileptonic decay}
\author{Steven Gottlieb \address[IU]{Indiana University, Bloomington, 
IN 47405, USA; Theory Group MS106, Fermilab, PO Box 500, Batavia, IL 60510, USA.} 
and Sonali Tamhankar\addressmark[IU]\thanks{Presented 
by S. Tamhankar}}
\begin{document}

\begin{abstract}
We present results from a lattice study of the semileptonic decay
$\Lambda_b \rightarrow \Lambda_c l \nu_l $.  We use $O(a^2, \alpha_s a^2)$
improved quenched lattices of the MILC collaboration, with lattice spacing
$\sim 0.13$ fm. For the valence quarks, the tadpole-improved clover action is
used, with the Fermilab method employed for the heavy quarks. Form factors
are extracted from the vector as well as the axial-vector part of the current.
\vspace{1pc}
\end{abstract}

\maketitle

\section{INTRODUCTION}
Current knowledge of the CKM matrix element $V_{cb}$ is derived from the
mesonic decays
$B \rightarrow \bar{D}^{*}l\nu$ or $B \rightarrow \bar{D}l\nu$.
Experimental knowledge of the $\Lambda_b$ semileptonic decay
can lead
to an independent estimate of $V_{cb}$ if the effect of
the strong interaction in the decay are understood, {\it e.g.}, via
lattice QCD.
A first lattice study of the baryonic semileptonic decay was
performed by the UKQCD collaboration~\cite{ukqcd}. We report
our initial results for the dominant form factors of this decay.

The semileptonic decay $\Lambda_Q \rightarrow \Lambda_{Q'} l\nu$ can be
parametrized in terms of six form factors, $F_i$ and $G_i$, for i = 1, 2, 3.
\begin{eqnarray}
\langle \Lambda_Q^{(s)}(v)|\mathcal{J}_{\mu}\hspace{-1cm}&&|\Lambda_{Q'}^{(r)}(v')\rangle =
\bar{u}^{(s)}_Q(v)  [  \gamma_{\mu}(F_1-\gamma_5 G_1)\nonumber\\ &+&
  v_{\mu} (F_2-\gamma_5 G_2) \nonumber\\
   &+& v'_{\mu}(F_3-\gamma_5
  G_3)]u^{(r)}_{Q'}(v'). \label{velform}
\end{eqnarray}
Here $\mathcal{J}_{\mu}$ is the weak current and $r,s$ are polarisation
states of the baryons.
Since both $\Lambda_b$ and $\Lambda_c$ are hadrons containing a single
heavy quark, heavy quark effective theory (HQET) is applicable~\cite{MW}.
 Hence the matrix element is taken between baryons of a given velocity,
and the form factors are functions of the scalar $\omega = v \cdot v'$.
To leading order in HQET, the combinations $F_1+v_0F_2+v_0'F_3$ and $G_1$
involving the dominant form factors $F_1$ and $G_1$ can be written
in terms of a single function, called the (baryonic) Isgur-Wise 
function, $\xi(\omega)$. This function is normalised at zero-recoil,
$\xi(1)=1$.

\section{SIMULATION PARAMETERS}
The simulations are performed on the Asqtad quenched lattices
at $\beta = 8.00$ generated by the MILC collaboration~\cite{asqtad}. These
are $\mathcal{O}(\alpha_sa^2)$ improved 
$20^3 \times$ 64 lattices, with $a^{-1}$ = 1.33 GeV, as determined from
$m_{\rho}$. We use 
three light quark masses near the strange quark mass, $\kappa_l =
0.1343, 0.1333, 0.1323$. Two heavy quark $\kappa$ values, 0.104 and 
0.114 bracket the charm quark, and other two, 0.064 and 0.077
bracket the bottom. We use the clover action for the valence quarks,
with a tadpole improved clover coefficient. The value for the tadpole
improvement factor $u_0$ is taken from the Landau gauge fixed mean link.
Fermilab formalism is used for the heavy quarks. Results are presented
for 300 lattices for two-point functions, and 237 lattices for three-point
functions.

\section{TWO-POINT RESULTS}
\begin{figure}\begin{center}
\includegraphics[width=7cm]{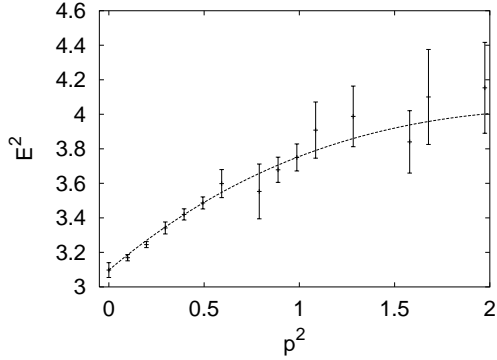}
\vspace{-1cm}
\caption{Dispersion relation for $\kappa_h$=0.114, $\kappa_l$ = 0.1323/0.1323.
The $E$ here is $E_1$, the parameter obtained from exponential
fits. The line shows the lattice dispersion relation.}
\vspace{-1cm}
\label{disp_fig}
\end{center}\end{figure}

\begin{figure}\begin{center}
\vspace{-1cm}
\includegraphics[width=7cm]{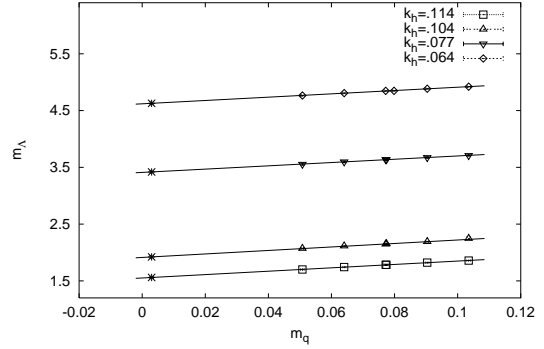}
\vspace{-1cm}
\caption{Chiral extrapolations of the measured heavy baryon masses to the
$u$ quark. We have used the light quark kinetic mass $m_2$ for the fit.}
\label{chiral_fig}
\vspace{-1cm}
\end{center}\end{figure}

\begin{figure}\begin{center}
\includegraphics[width=7cm]{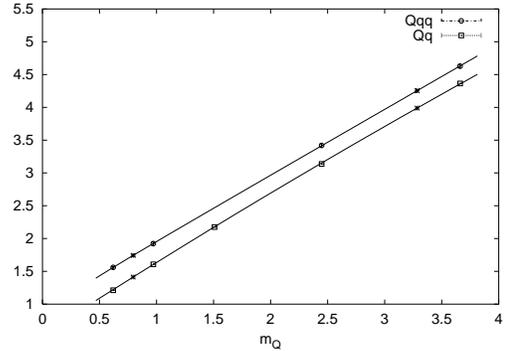}
\vspace{-1cm}
\caption{The heavy baryon mass, plotted as function of the
heavy quark mass. Also shown are the heavy-light
meson masses, taken from studies of the MILC collaboration.
The bursts correspond to the b and c quark.}
\vspace{-1cm}
\label{baryon_mass}
\end{center}\end{figure}

The dispersion relation is shown in Fig.~\ref{disp_fig}. The fitted energy values
agree very well with the expectation from the lattice dispersion relation.
The chiral extrapolations for a fixed heavy quark mass are shown in 
Fig.~\ref{chiral_fig}. The baryon kinetic mass $M_2$ is estimated
as $M_2=M_1+m_2-m_1$, Where $M_{2(1)}$ and $m_{2(1)}$ are the baryon 
and heavy quark kinetic(rest) masses respectively.
We use a linear fit for these extrapolations. 
In Fig.~\ref{baryon_mass}, we have shown the chirally extrapolated baryon mass 
as a function of the heavy quark mass, along with the coresponding 
meson masses taken from the MILC collaboration. Our values for 
$m_{\Lambda_b}$ and $m_{\Lambda_c}$ are 5.626(36)GeV and 2.300(27)GeV.  

\section{THREE-POINT RESULTS}
Different form factors contribute to different matrix elements
in the three-point function. For $\mu = 0$, the dominant contribution
to three-point functions comes from the vector form factors and
for $\mu = i$, axial-vector form factor $G_1$ gives the dominant contribution.
We present results for the Isgur-Wise function from vector as well as
axial-vector data. $\Lambda_b$ is created at time 0 and $\Lambda_c$ is 
annihilated at time $t_x \equiv 16$ in lattice units. 
The time at which the current acts is varied, and we study 
three-point function as a function of this time $t_y \equiv t$.
For the results presented here, the intial baryon is at rest and the final
baryon is moving with different velocities giving different values for $\omega$.
On the lattice, one is restricted to region near $\omega=1$ as data starts
getting noisy for high momenta. In this region, $v'_0$ can be approximated
by 1.
Then for an initial baryon of mass $M'$ decaying to a final baryon of mass $M$ moving with a momentum $\vec{q}$, if we consider the sum of the 
co-efficients of $I$ and $\gamma_0$, for large $t_y$ and $t_x-t_y$,
the three-point expression simplifies to
\begin{figure}\begin{center}
\includegraphics[width=7cm]{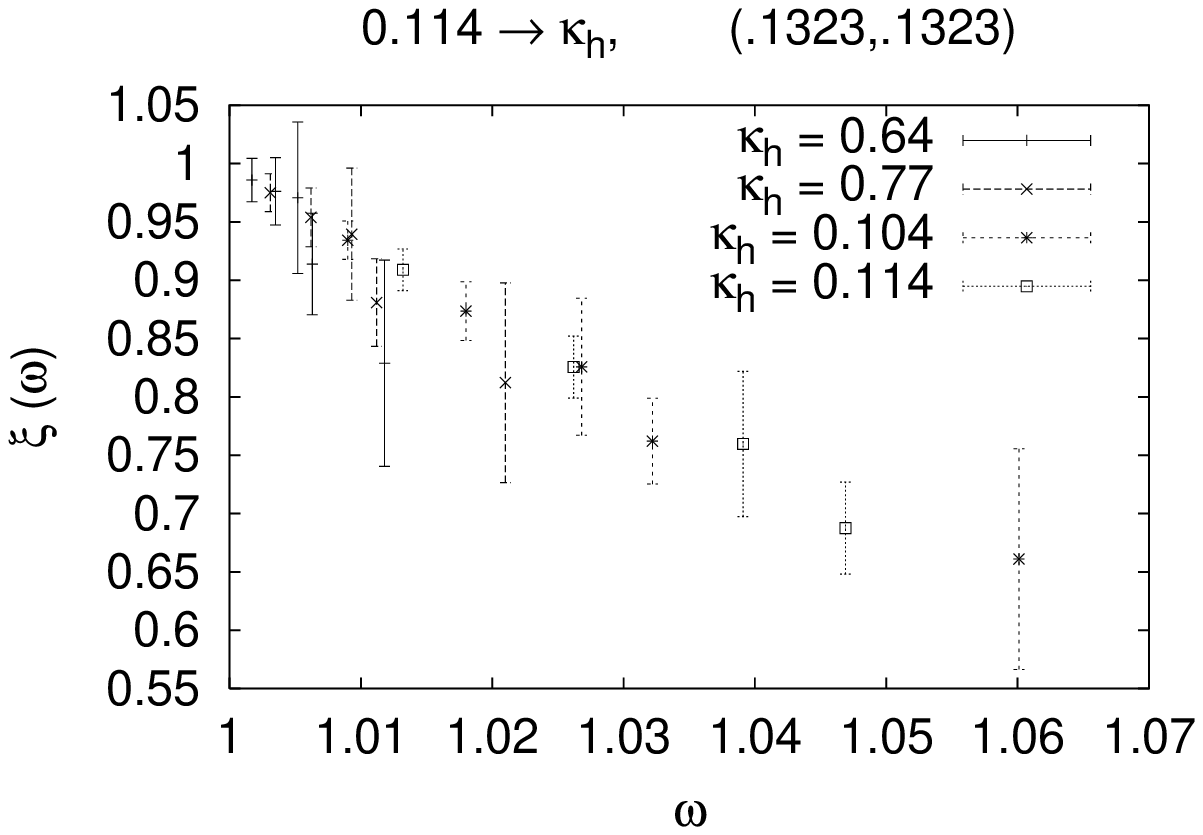} \\
\vspace{-1cm}
\caption{Isgur-Wise function from the vector current. $\kappa_{h1}$ is
0.114 for all these points, and the points corresponding to four
different $\kappa_{h}$ are shown with four different symbols.
}
\vspace{-1cm}
\label{vec_IW}
\end{center}\end{figure}
\begin{figure}\begin{center}
\includegraphics[width=7cm]{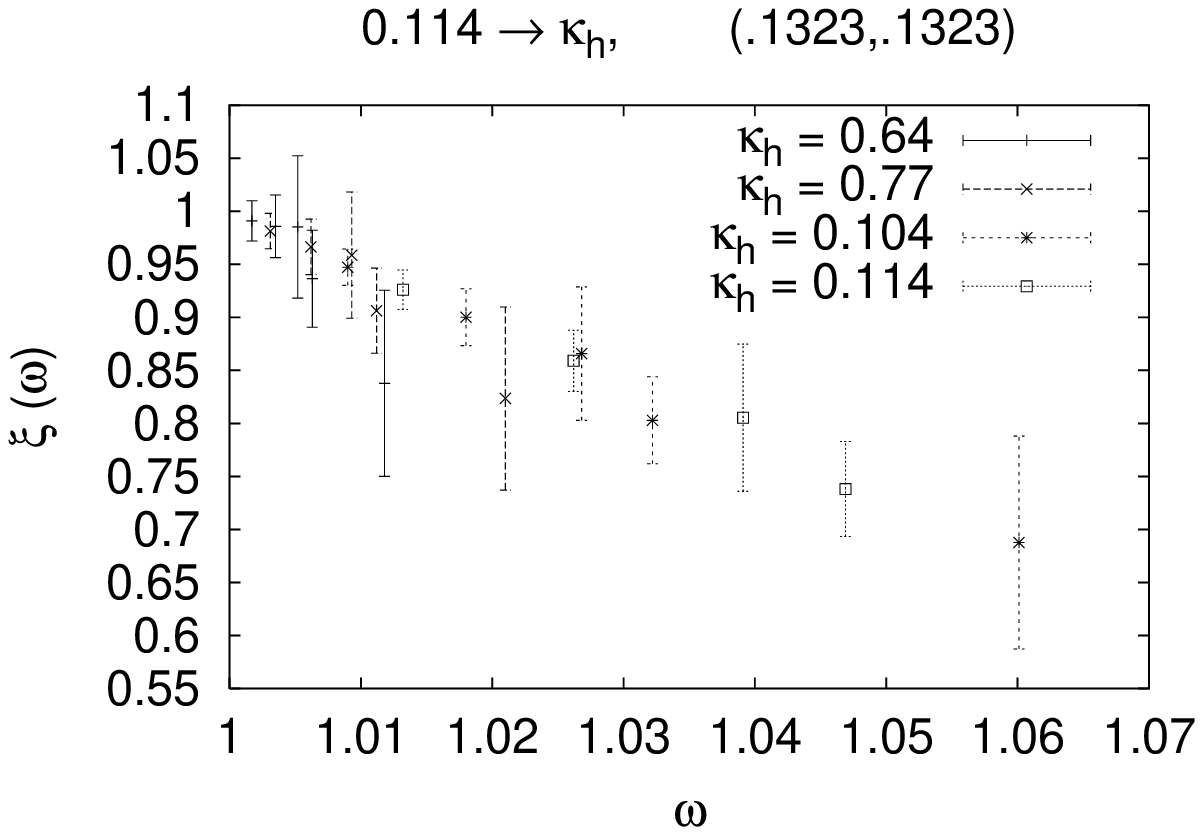} \\
\vspace{-1cm}
\caption{Isgur-Wise function from the axial-vector current. As before, $\kappa_{h1}$
is 0.114 for all these points, and the points corresponding to four
different $\kappa_{h}$ are shown with four different symbols.
}
\vspace{-1cm}
\label{ax_IW}
\end{center}\end{figure}
\begin{eqnarray}
C(t_y)&&\hspace{-0.7cm}=\frac{Z_l Z'_s(|\vec{q}|) }{16M'E}e^{-t_xM'}e^{(M'-E)t_y}
        4M'\nonumber\\&&\hspace{-0.7cm}(F_1(\omega)+F_2(\omega)+F_3(\omega))2(E+M),
\end{eqnarray}
where $Z_l$ and $Z'_s$ are known from the two-point functions.
We fit this to a form $Ae^{-Bt}$ and consider the ratio
\begin{eqnarray}
\frac{A[(M',\vec{0}) \rightarrow (M,\vec{q})]}{A[(M',\vec{0}) \rightarrow (M,\vec{0})]} &&\hspace{-0.5cm}=
\Big( \frac{F_1(\omega)+F_2(\omega)+F_3(\omega)}{F_1(1)+F_2(1)+F_3(1)} \Big)\nonumber\\
&&\hspace{-0.9cm}\cdot\Big( \frac{Z'_s(|\vec{q}|)}{Z'_s(0)}\Big) \cdot \Big( \frac{E+M}{2E}\Big).
\end{eqnarray}
First factor on the RHS is\footnote{We follow Ref.~\cite{ukqcd} in 
using this definition. This way of defining the Isgur-Wise
function agrees with the conventional definition up to $\mathcal{O}$(1/m) corrections,
which are further multiplied by (1-$\omega$). For the small momenta accessible
on the lattice, $\hat{\xi}$ is a very good approximation to the baryonic Isgur-Wise
function. The suffix $QQ'$ emphasizes that the infinite mass limit has not been taken.} the Isgur-Wise function $\hat{\xi}_{QQ'}$.
The second and third factors are known from the two-point functions.
The third factor may be approximated by 1 to 0.5 per cent accuracy.
The second factor differs from 1 by upto 10\% over our range of $\vec q$.
The ratio is independent of the renormalization constant $Z_V$ 
because we have the same heavy quark transition
in both numerator and denominator.

Our results for the Isgur-Wise function from the vector current
are shown in Fig.~\ref{vec_IW}. The Isgur-Wise function obtainted
from the axial-vector current ($\mu = i$ case) is shown in Fig.~\ref{ax_IW}.
The Isgur-Wise function seems to be quite insensitive to the heavy
quark mass.

\begin{figure}\begin{center}
\includegraphics[width=7cm]{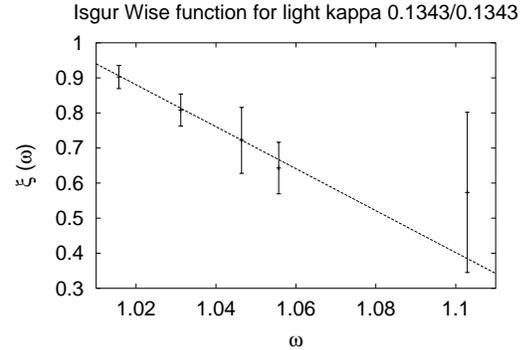}
\includegraphics[width=7cm]{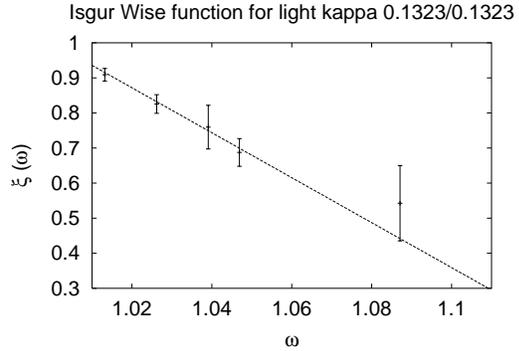}
\vspace{-1.1cm}
\caption{Isgur-Wise function for $\kappa_l=0.1343/0.1343$, and
$\kappa_l=0.1323/0.1323$, our highest and lowest values for $\kappa_l$.}
\label{light}
\vspace{-1.3cm}
\end{center}\end{figure}

We have also studied the light quark mass dependence of the Isgur-Wise function.
The Isgur-Wise function is expected to fall slower for smaller light quark
masses, by a heuristic argument. We do see such a trend in this preliminary
study, but it is very far 
from clear with the statistical errors we have. This is shown in Fig.~\ref{light}.

The calculations were done on the IBM SP at Indiana University.
We gratefully acknowledge the hospitality
of the Fermilab Theory Group.

\end{document}